# GANDA: A Deep Generative Adversarial Network Conditionally Generates Intratumoral Nanoparticles Distribution Pixels-to-Pixels

*Jiulou Zhang#, Yuxia Tang#, Shouju Wang\**

Department of Radiology, The First Affiliated Hospital of Nanjing Medical University, Nanjing, Jiangsu, China
E-mail: shouju.wang@gmail.com

**Abstract:** Intratumoral nanoparticles (NPs) distribution is critical for the success of nanomedicine in imaging and treatment, but computational models to describe the NPs distribution remain unavailable due to the complex tumor-nano interactions. Here, we develop a Generative Adversarial Network for Distribution Analysis (GANDA) to describe and conditionally generates the intratumoral quantum dots (QDs) distribution after *i.v.* injection. This deep generative model is trained automatically by 27 775 patches of tumor vessels and cell nuclei decomposed from whole-slide images of 4T1 breast cancer sections. The GANDA model can conditionally generate images of intratumoral QDs distribution under the constraint of given tumor vessels and cell nuclei channels with the same spatial resolution (pixels-to-pixels), minimal loss (mean squared error, MSE = 1.871) and excellent reliability (intraclass correlation, ICC = 0.94). Quantitative analysis of QDs extravasation distance (ICC = 0.95) and subarea distribution (ICC = 0.99) is allowed on the generated images without knowing the real QDs distribution. We believe this deep generative model may provide opportunities to investigate how influencing factors affect NPs distribution in individual tumors and guide nanomedicine optimization for molecular imaging and personalized treatment.

## 1. Introduction

Nanoparticles (NPs) are still a powerful and promising tool for the diagnosis and treatment of tumors. The NPs are believed to accumulate in solid tumors more selectively because they can leak out from the neovessels and stay in tumors for a longer time, which is called the enhanced permeability and retention (EPR) effect [1]. However, recent studies have increasingly realized that the EPR effect strongly varies among individuals [2]. The heterogeneity of the tumor microenvironment, such as blood vessel density, cell density and stroma components, highly affects the delivery and distribution of NPs across tumors [3]. So it is clear that the "one-size-fits-all" design is unrealistic. How to deepen our understanding of the relationship between intratumoral NPs distribution and tumor microenvironment is critical for the individualized design of NPs.

Modeling the intratumoral NPs distribution is an important but challenging task due to the complex tumor-nano interaction [4]. Firstly, the tumor microenvironment imparts numerous pathophysiological barriers to NPs delivery. The NPs extravasation was affected by the blood flow, the fluid drainage, the interstitial fluid pressure, the cell density and the extracellular matrix in the tumors [3]. Secondly, the NPs distribution also highly relies on NPs characteristics, such as size, shape, circulatory half-time, charge and surface modification [5,6]. Finally, the EPR effect is affected by concomitant therapies. Several EPR-enhancer has been proved to change the NPs distribution, including vascular mediators, radiotherapy, photodynamic therapy, hyperthermia and sonoporation [7]. Such convoluted and high-dimensional impact on the intratumoral NPs distribution makes it difficult to be described by traditional hypothesis-based computational models.

Deep generative models, such as generative adversarial networks (GAN) and variational autoencoders (VAE), are the state-of-the-art methods to estimate data distribution [8,9]. These models are trained by large-scale datasets and can not only learn the features of data distribution but also generate new data points as similar as the real data. Very recently, these models have been introduced to the field of bioimaging and bioinformatics to augment small datasets [10,11]. Some studies showed the potential of using the generated data for downstream analysis, such as image segmentation [12] and human embryo cell stage recognition [13]. However, no reports explored the potential of using deep generative models to describe the interaction between NPs and tumors *in vivo*, such as analyzing the intratumoral NPs distribution after *i.v.* injection.

Here, we proposed a deep generative model to conditionally generate intratumoral quantum dots (QDs) distribution pixels-to-pixels under the constraint of given tumor vessels and cell nuclei information. This Generative Adversarial Network for Distribution Analysis (GANDA) was trained by the whole-slide images of 4T1 breast cancer sections. We found the GANDA generated intratumoral QDs distribution is very close to the real distribution. Virtual quantitative analysis, such as QDs extravasation distance calculation and subarea distribution estimation, could be performed on the generated intratumoral QDs distribution with high accuracy and reliability.

## 2. Results

### 2.1. Highly heterogeneous distribution of NPs.

To visualize the NPs distribution, we intravenously injected 20-nm PEGylated CdSe/ZnS quantum dots (QDs) into six 4T1 mice models of breast cancer (Table S1). The characterization of QDs is shown in SI Appendix, Figure S1 and Table S2. The blood half-life time of QDs was 2.42 h (Figure S2). QDs were selected as model NPs because their fluorescence bleaches minimally and can be readily detected in tumor sections. The tumor blood vessels were labeled with a primary CD31 antibody and a secondary antibody conjugated with Alex-Fluor-488. The tumor cell nuclei were identified with DAPI staining. **Figure 1A** shows the whole-slide images of the training group (tumor No. 1 to No. 5). The images consist of three channels, which contain spatial information of tumor cell nuclei, vessels and QDs, respectively. The mean intratumoral QD intensity was related to the total amount of QDs in tumor ($R^2$=0.91, Figure S3).

To assess the heterogeneity of QDs distribution, we used QuPath to analyze the whole-slide images of tumor sections. The cell, vessel and QD densities were estimated as the area ratios of DAPI-, Alex-Fluor-488- and QD- positive regions to the whole section or the selected regions of interest (ROIs). The average cell, vessel and QD densities of these five tumors were 38.2 ± 8.4%, 11.0 ± 7.1‰ and 7.7 ± 3.2‰. It is noted that the vessel density varied sevenfold, and the QD density varied fourfold among these tumors. When we look at the ROIs in each tumor, the vessel and QD densities exhibited even greater heterogeneity (Figure 1B). For tumor No. 1, the vessel densities of ROIs ranged from 1.1‰ to 108.7‰, and the QD densities of ROIs ranged from 0.0‰ to 77.2‰. No discernible relationship was found among the cell, vessel and QD densities of ROIs (Figure 1C). In summary, these data show that the QDs distribution is highly heterogeneous both intra- and inter-tumors and has a complicated relationship with tumor characteristics like cell and vessel density.

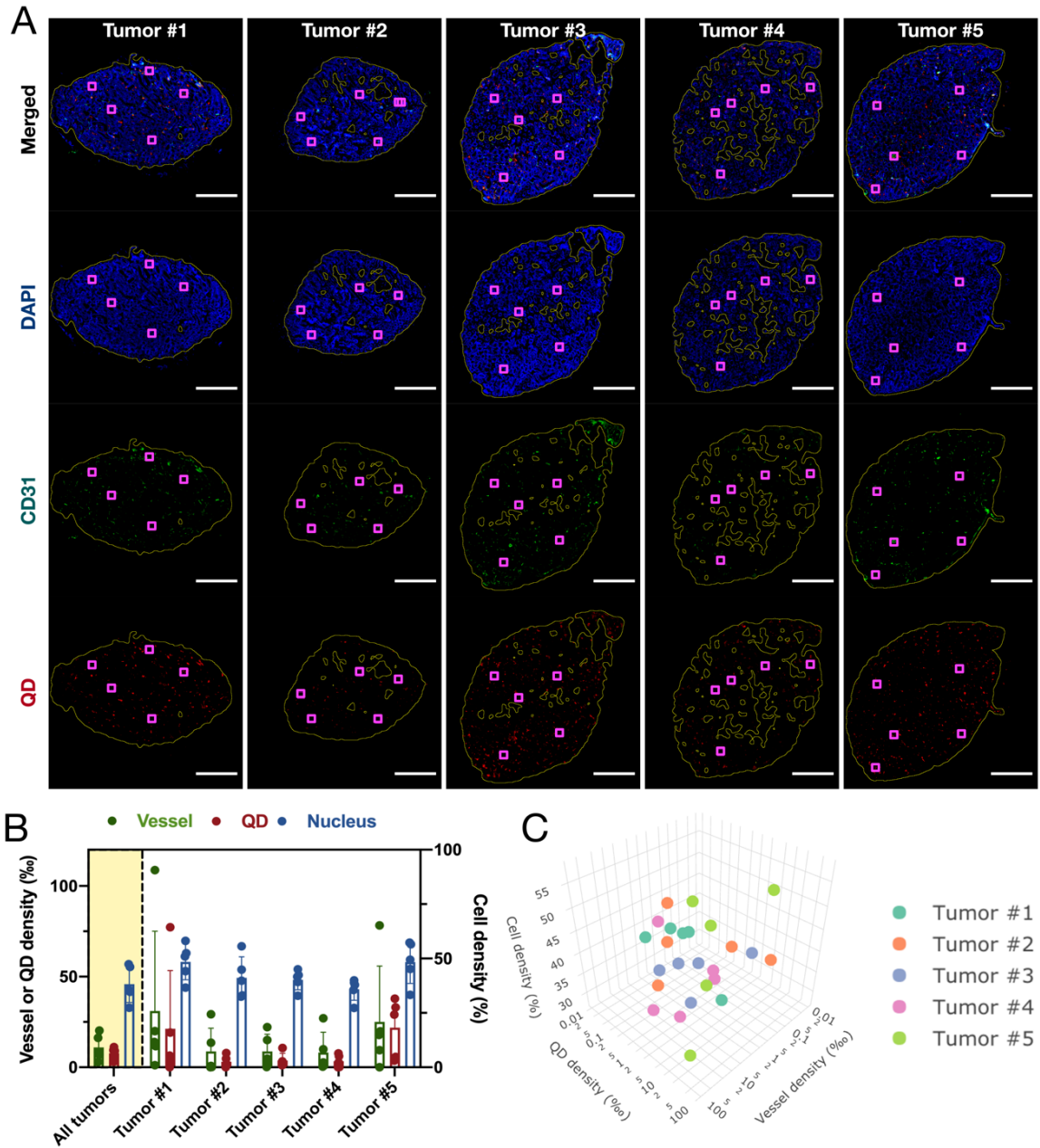

**Figure 1.** The heterogeneous tumor characteristics and QDs distribution. (A) Whole-slide images of frozen tumor sections, composed of DAPI (Blue: cell nuclei), CD31 (Green: vessels) and QD (Red: PEGylated CdSe/ZnS quantum dots). Scale bar, 2 mm. Yellow lines indicate the whole tumor edges. Magenta squares indicate the ROIs. (B) Cell, vessel and QD densities of the five whole tumors (yellow part) and ROIs of each tumor. Error bars show SD. (C) The cell, vessel and QD densities of ROIs of each tumor in a 3D plot.

## 2.2. The architecture of GANDA.

To investigate the relationship between QDs distribution and tumor characteristics, we established a deep generative model called GANDA. The workflow of training and testing the model is depicted in **Figure 2**. The whole-slide images of the five training group tumors were decomposed to 512 pixels × 512 pixels patches (27775 patches in total). A generator network based on deep fully convolutional networks (FCN) was trained to generate QD-channel patches under the condition of corresponding DAPI or/and CD31 channels. A discriminator network was trained simultaneously to identify the generated QD-channel patches are real or fake. The adversarial and pixel-wise loss was propagated backward to enhance the generator until the discriminator can not distinguish the generated QD-channel patches from real ones anymore. The architecture of the generator and the discriminator was detailed in SI Appendix, Figure S4 and S5. The adversarial loss (generator and discriminator loss) and pixel loss during training were recorded in Figure S6 to show the convergence of the model.

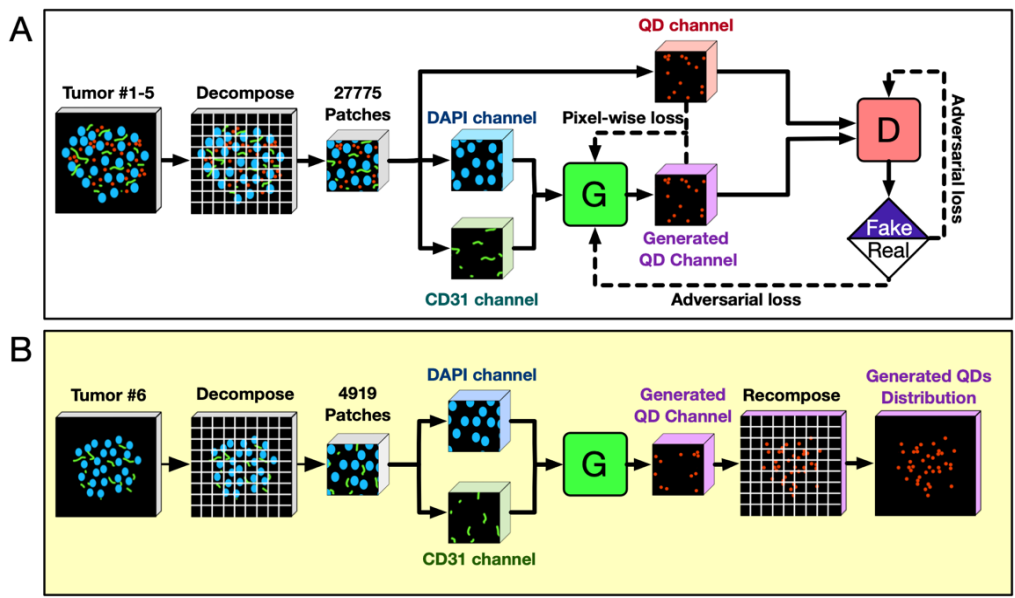

**Figure 2**. The framework of GANDA. (A) in the training phase, the whole-slide 4T1 tumor section images are decomposed into patches. The DAPI and/or CD31 channels of tumor No.1 to No. 5 are learned by the generator network (G) to generate the corresponding QD channel. A discriminator network (D) is trained to distinguish between the generated and real QD channels. The adversary and pixel-wise losses propagate backward to upgrade the generator until the discriminator could not distinguish the generated and real QD channels patches anymore. (B) in the testing phase, the trained generator conditionally generates the QD-channel patches according to the DAPI and/or CD 31 channel patches of tumor No. 6. The generated patches are recomposed to the image of intratumoral QDs distribution.

To test the reliability of GANDA, the DAPI and/or CD31 channels of tumor No. 6 were decomposed to 512 pixel × 512 pixels patches (4919 in total) and input to the trained generator to generate corresponding QD-channel patches. The generated patches were recomposed to a full image of intratumoral QDs distribution with the same spatial resolution of the original DAPI and CD31 channels (*i.e.*, the generated intratumoral QDs distribution is a 1:1 pixels-to-pixels mapping from the images of tumor cell nuclei and/or vessels).

## 2.3. Impact of source channels on the generated intratumoral QDs distribution.

To determine the impact of source channels on the training of GANDA, the generated intratumoral QDs distribution was compared to the real distribution. **Figure 3** shows that the GANDA model trained by CD31 and DAPI + CD31 channels generates realistic QDs distributions. Noted that the GANDA trained by the DAPI + CD31 channel generated the sharpest image with the least residual error. In contrast, the GANDA learned from the CD31 channel generated a more blurred image. The mean squared error (MSE) between the generated and real QDs distribution was also smallest for the GANDA trained by DAPI + CD31 channel (MSE = 1.871), comparing with GANDA trained by DAPI (MSE = 72.321) or CD31 (MSE = 2.616) channel only. The pixel loss of GANDA trained by DAPI + CD31 channel is also smaller than that trained by DAPI only (Figure S7). These data demonstrated the GANDA learned most of the features of intratumoral QDs distribution from the tumor vessels information and adding the cell nuclei information can further improve its accuracy.

To show the stability of model training, we trained GANDA nine times with different initializations, and recorded the pixel loss of each model. It is shown that all models converged after 50 epoches of training (Figure S8). The average pixel loss was $8.68 \times 10^{-5} \pm 5.73 \times 10^{-6}$.

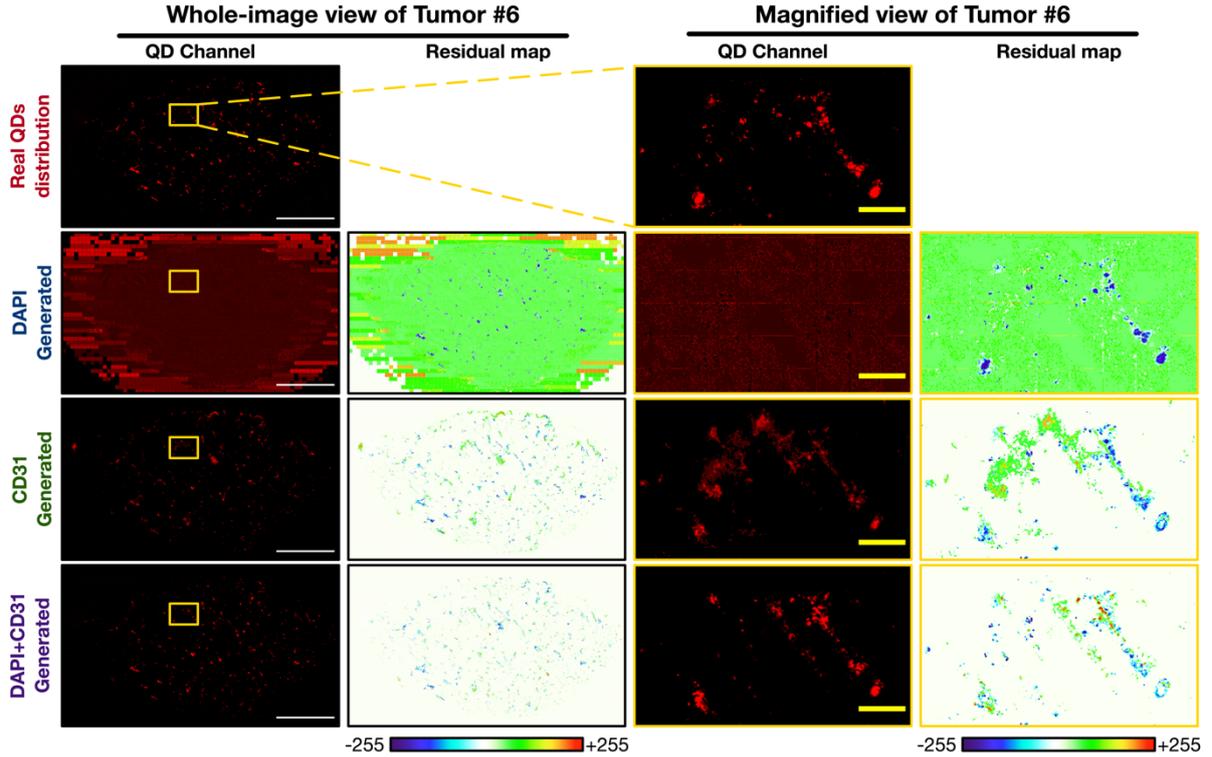

**Figure 3**. The whole-image and magnified views of the real and generated intratumoral QDs distributions and their corresponding residual maps. The top row is the real QDs distribution. The 2nd and 3rd rows are the generated QDs distributions from GANDA trained by DAPI or CD31 channel. The 4th row is the generated QD distribution from GANDA trained by DAPI + CD31 channel. White scale bars, 2 mm; Yellow scale bars, 200 μm.

Mode collapse is a major issue for training GAN. The histograms of the average QDs fluorescence intensity were depicted to show the diversity of original and generated patches of tumor No. 6, as shown in Figure S9. The patches with an average intensity equal to zero were excluded. The generated data showed a similar left-skewed distribution to the original data. It is noted the intensity distribution of generated data is slightly shifted to the left because the very weak signal in original patches was omitted by the model as background.

**2.4. Virtual quantitative analysis based on the generated intratumoral QDs distribution.**
To show the potential of GANDA in downstream analysis, we measured the QD densities of the generated and real QD channels of tumor No. 6. As shown in **Figure 4A** and 4B, the measured QD densities of the generated image are very close to those of the real one. There is a significant linear relationship between the measured QD densities of ROIs in generated and real images (slope = 1.07; $R^2$ = 0.93; p < 0.00001). The reliability of the GANDA estimation is evaluated by intraclass correlation (ICC) and the result is excellent (ICC = 0.94; 95% confidence interval (CI) = [0.86, 0.98]). It indicates the GANDA can explain 94% of the heterogeneity of intratumoral QDs distribution, and its estimation of QD density would be no more than 7% higher than the real QD density.

Because the generated QD channel has the same size and spatial resolution as DAPI and CD 31 channels, we can perfectly merge these channels to generate a triple-channel whole-slide image. The obtained images can be used for virtual downstream analysis, such as extravasation distances calculation and subarea distribution analysis. Figure 4C and Table S3 show that the measured QDs-to-Vessels distances calculated from the generated image (median distance: 10.94 μm) are very close to that from the real image (median distance: 11.64 μm). The reliability of QDs-to-Vessels distances estimation is almost perfect (ICC = 0.99; 95% CI = [0.99, 1.00]). To test the ability to perform subarea distribution analysis on generated intratumoral QDs distribution, tumor No.6 was divided into five subareas from the inside out as shown in Figure 4D and 4E. The relationship among cell, vessel, and QD density was depicted in Figure S10. The QD density showed moderate correlation with cell density (p = 0.04, $R^2$ = 0.79). The measured subarea QD densities from the GANDA generated image also exhibited great accuracy (slope = 1.08, $R^2$ = 0.99, p < 0.00001) and reliability (ICC=0.95, 95% CI = [0.74, 0.99]). These results demonstrated the potential of using GANDA to quantitatively analyze intratumoral QDs distribution without knowing their real distribution.

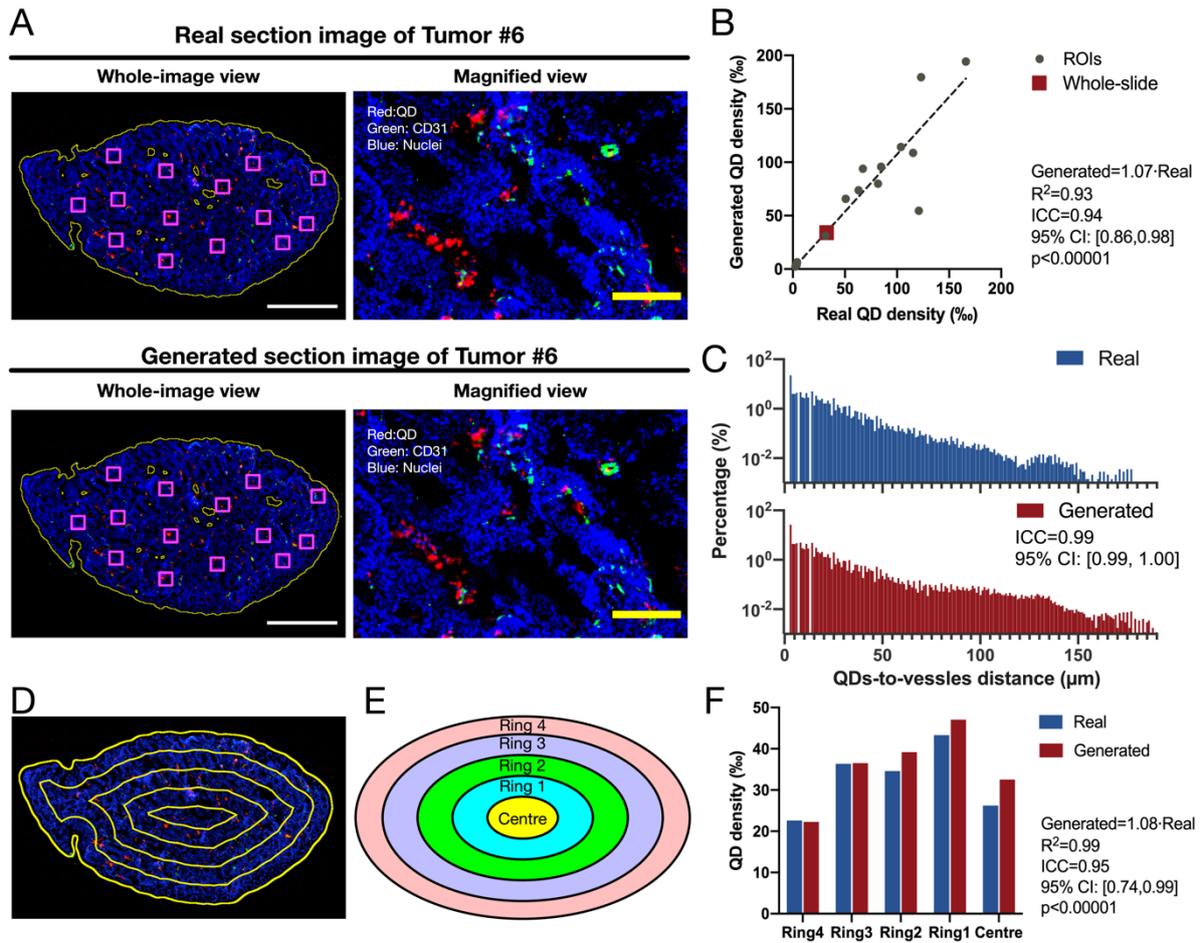

**Figure 4**. The virtual quantitative analysis based on the generated intratumoral QDs distribution of tumor No. 6. (A) The whole-slide and magnified views of real and generated section images. White scale bars, 2 mm. Yellow scale bars, 200 μm. Yellow lines indicate the whole tumor edges. Magenta squares indicate the randomly selected ROIs. (B) The scatter plot of the measured QD densities of the whole tumor and ROIs on generated and real section images. The QD density of the whole tumor No. 6 is indicated by a red square. (C) The QDs-to-vessels distance distribution calculated from the real and generated section images. (D) and (E) indicates the division of subareas of tumor No. 6. The width of each ring is 550 μm. (F) The measured QDs density of each subarea on generated and real section images of tumor No.6.

2.5 The generalization capability of GANDA

To show the generalization capability of GANDA, the model was evaluated by cross-validation and tested in another B16 tumor model. For the cross-validation, the images of tumor No. 1-4 and 6 were used as a training dataset to generate the QDs distribution of No. 5. As expected, the generated distribution was very close to the real distribution (MSE = 2.736, Figure S11).

Furthermore, the plot of validation loss showed no increase at the end of the training (Figure S12), indicating the model was not over-fitted.

For the test in B16 model, the real and generated QDs distribution was depicted in Figure 5. It is noted that the generated QDs distribution was similar to the real distribution (MSE = 7.464) with an increased noise signal. The QD densities were measured from five randomly selected ROIs, and a significant linear relationship was identified between the real and generated images of B16 (y = 0.52x+ 6.36, $R^2$ = 0.967). These results indicate that the GANDA model has robust applicability across different tumor models.

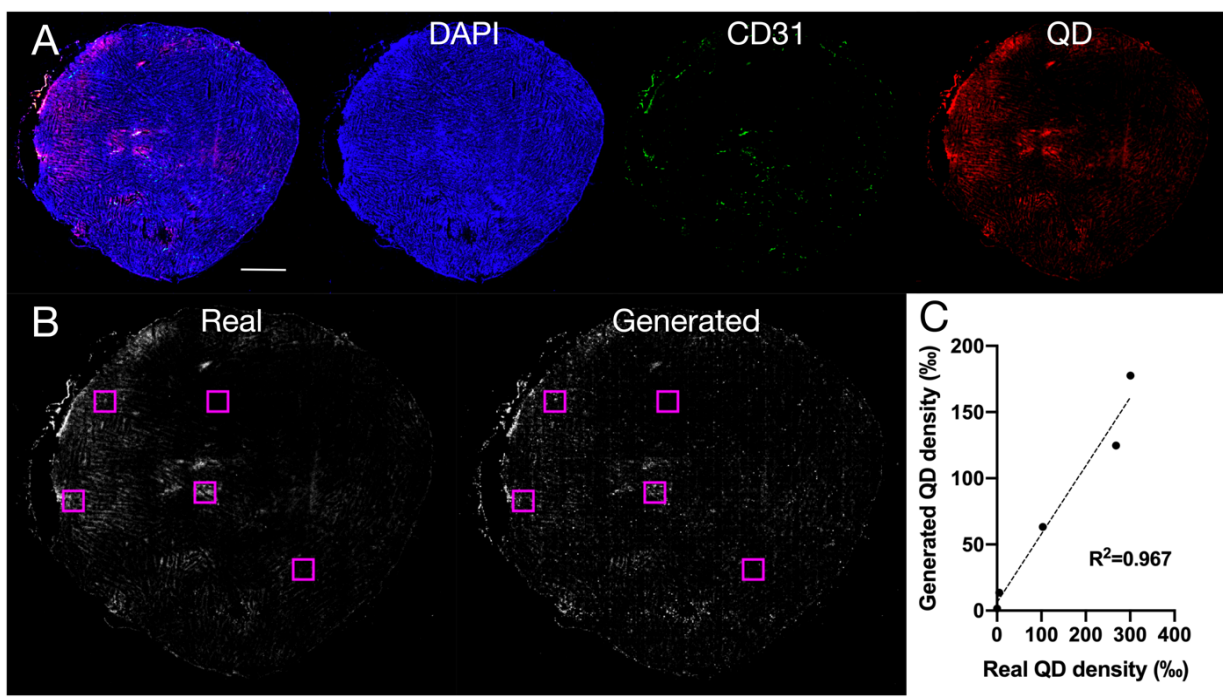

**Figure 5**. The test of GANDA in a B16 tumor model. (A) The whole-slide section image of B16 tumor. White scale bar, 1 mm. (B) The real and generated QDs distribution of B16 tumor in grayscale. Magenta squares indicate the randomly selected ROIs. (C) The scatter plot of the measured QD densities of the ROIs on generated and real section images.

## 3. Discussions

We established a clinically relevant 4T1 model in immune-competent mice, which is widely used to study drug and nanoparticle delivery as well as tumor heterogeneity [14,15]. Given the heterogeneity of cell nuclei and vessels observed in this 4T1 breast cancer model, it is difficult to describe the intratumoral NPs distribution precisely [16]. Here, as a proof-of-concept, we built a deep generative model called GANDA to describe and conditionally generate QDs distribution pixels-to-pixels according to the section images of tumor vessel and cell nuclei. We showed that the generated intratumoral QDs distribution by GANDA has high accuracy and reliability and could be used for virtual quantitative downstream analysis, such as extravasation distance calculation and subarea distribution estimation, even without knowing the real QDs distribution. Considering the tumor microenvironment heterogeneity, optimizing the intratumoral distribution and maximizing drug access are pivotal to the success of nanomedicine[17]. Therefore, the virtual quantitative analysis of intratumoral NPs distribution in a clinically relevant tumor model is important for the design of nanomedicine.

The GANDA model is based on deep convolutional generative adversary network (DCGAN), which is the driving force of deep learning and computer vision [18–20]. The key advantage of deep learning is that it can automatically learn features, while conventional machine learning requires considerable expertise to design feature extractors [8,21,22]. As shown in our report, the GANDA automatically learned the tumor features from the images of tumor cell nuclei and vessels. In contrast, one very recent study based on machine learning methods (support vector machine) has to manually extract features of NPs and tumors to predict the NPs delivery to micro-metastases [23].

Another advantage of GANDA is it preserved the spatial and contextual information of tumor sections by using FCN as the generator. The FCN architecture replaced the fully connected layers with upsampling operators; hence, it can keep the output resolution the same as the input [18,24]. In our work, the GANDA generated QD channels with the same resolution as DAPI and CD31 channels and allowed us to analyze the spatial relationship between QDs and vessels. This is not achieved in previous work based on conventional machine learning methods, because the spatial information of NPs distribution was lost in the pooling or fully connected layers of the networks [23,25,26].

One of the biggest challenges of applying artificial intelligence approaches (including deep learning and machine learning methods) to analyze nano-bio interactions is prohibiting large amounts of data required to train the models [27,28]. Measuring the characteristics of tumors and NPs hundreds and thousands of times is extremely costly and time-consuming. For example, one very recent report used at least 30 min to label NPs, tumor vessels and cell nuclei for one tumor section [23]. A training dataset containing 1000 tumor sections would take more than 500 hours just for image labeling, not to mention the time and labor required to prepare these sections. In our report, to use the whole-slide images more efficiently, we decomposed these images into small patches. In this way, we obtained 27775 patches from only five tumor sections. Furthermore, thanks to the automatically learning ability of GANDA, we do not need to label the NPs, cell nuclei and vessels manually.

Our results indicated that the features extracted from the tumor vessels are predictive of the intratumoral NPs distribution. This observation is consistent with the previous belief that the tumor vessels are the foundation of the NPs accumulation in tumors. However, it should be noted that GANDA does not specify what mechanism (EPR effect, trans-endothelial pathways, *etc*.) dominates the extravasation of NPs [29]. One possible pitfall in interpreting our results is that the tumor cell density is irrelevant to the NPs distribution. However, the limited contribution from the DAPI channel to the predictive power of GANDA might come from the relatively small variation of cell density among our tumor models. In our 4T1 breast cancer model, the cell density varied only 1.6-fold, while the vessel density varied 7-fold among tumors. Some previous reports suggested that the high tumor cell density is crucial for the NPs diffusion in tumors because high tumor cell density reduced the interstitial space and increased the interstitial fluid pressure [30]. Therefore, we need more types of tumors to train the GANDA to determine the real impact of cell density on the NPs distribution.

Though the NPs are believed to accumulate in tumors selectively due to the EPR effect, the persistent unfilled gap between animal experiments and clinical translation has raised considerable argument over this notion [31–33]. The scarcity of robust computational methods capable of modeling NPs distribution makes it difficult to explain the reason for the gap [16,34]. Does the gap come from the heterogeneity of the EPR effect or the incorrect mechanism used to explain the NPs extravasation? Is it possible to enhance the EPR effect by personalized concomitant therapies? Is there any way to identify patients who would benefit from nanomedicine? We envision that the deep generative models like GANDA would be a powerful tool to help answer these questions. For example, the generated intratumoral NPs distribution from GANDA can be used as a "phantom" to study the effect of concomitant therapies on the NPs distribution in the same mouse and the same tumor. A workflow of the potential applications of GANDA for preclinical studies and the clinical translation is depicted in **Figure 6A** and 6B.

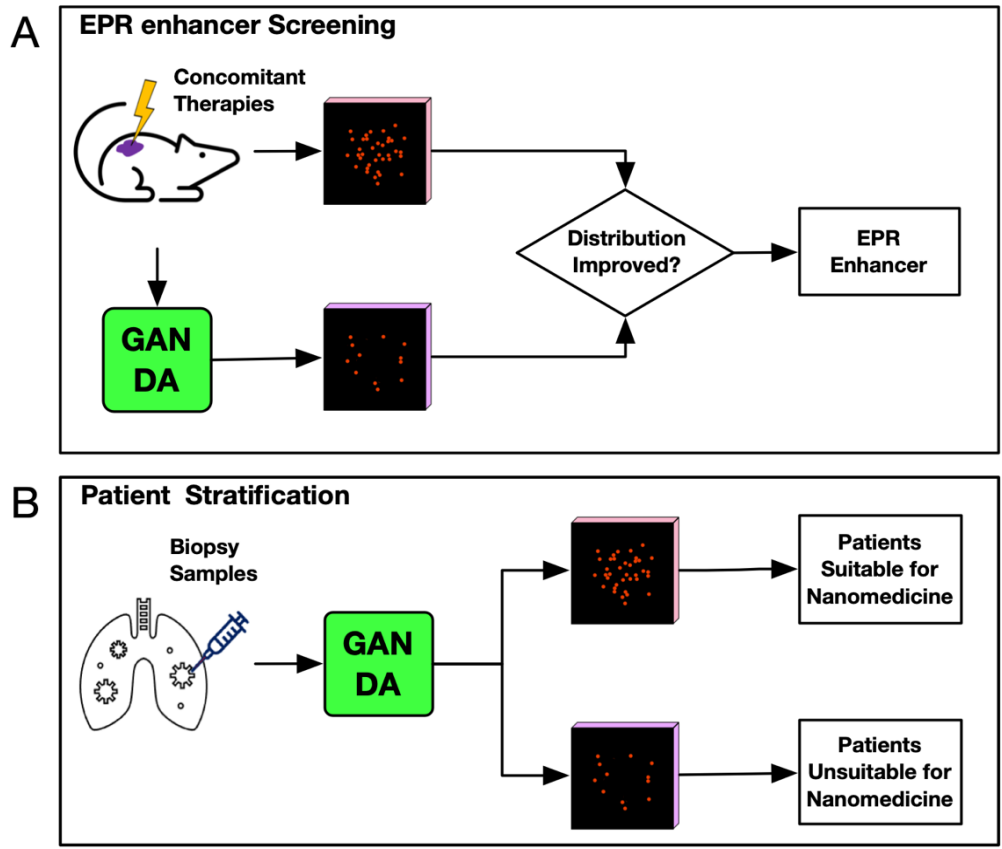

**Figure 6**. The potential applications of GANDA. (A) In the preclinical phase, the intratumoral NPs distribution generated by GANDA can be used as a phantom to precisely evaluate the effect of concomitant therapies on the NPs distribution and thus screen out EPR enhancers. (B) In the clinical trial phase, the GANDA can stratify patients based on biopsy samples and personalized the regime of nanomedicines for individuals.

As a pilot study, our GANDA model has several limitations. First, the training dataset only contains one tumor model. In the future, more types of tumor models, such as the orthotopic tumor mouse model, genetically engineered mouse models, and patient-derived xenograft mouse models, should be included to train the GANDA; Second, we only stained vessel endothelial cells and tumor cell nuclei in this study. Other tumor microenvironment components, like extracellular matrix and stromal cells, should also be stained and learned by GANDA; Thirdly, although the immunohistochemistry staining method is well established and readily available, it is not standardized across labs. Finally, the GANDA only predicts the NPs distribution in 2D images. However, the current method would be readily adopted for 3D spaces by combining 3D microscopy techniques (tissue clearing and light-sheet microscopy or micro-optical sectioning tomography) with 3D generative adversarial networks [10,35–37]. Despite these limitations, the present work demonstrated GANDA as a useful tool to predict the spatial distribution of NPs.

## 4. Conclusion

In summary, this study demonstrated the feasibility of using deep generative models to investigate the complex tumor-nano interactions with pixel-level accuracy and high reliability. It also provided GANDA to generate the intratumoral NPs distribution, which enables virtual quantitative downstream analysis.

## 5. Materials and Methods

*Characterization of QDs.* COOH-PEG-coated CdSe/ZnS QDs (304-CdSe/ZnS-PEG-COOH-620) were obtained from Hangzhou Fluo Nanotech company, China. The hydrodynamic size and zeta potential of QDs were measured using a PALS90 instrument (Brookhaven). The fluorescence excitation and emission spectra were obtained using a fluorescence spectrophotometer F-6400. The morphology of QDs was imaged using an FEI TF20 transmission electron microscopy.

*Breast cancer model.* Six tumors were induced by injecting $1 \times 10^6$ 4T1 cells into the right flank of BALB/c mice and allowed to grow for ten days until the tumor volume reaches about 100 mm$^3$. One hundred microliters of 2 nM QDs were injected intravenously. The mice were sacrificed 24 hours post-injection, and the harvested tumors were embedded in optimal cutting temperature (OCT) compound and frozen. All animal work was done in accordance with and approved by the Nanjing Medical University Animal Care Committee.

*In vivo circulation time of QDs.* Three mice were intravenously injected with QDs in PBS buffer (150 µL, 2 nmol/mL). At time points of 1, 2, 6, 12 and 24 h after injection, 20 µL blood were collected from the tail. The blood samples were then diluted to 200 µL and the emission intensity was measured at 622 nm by a microplate reader to determine the concentration of QDs from a standard curve. The calculation formula is %ID/g = The concentration of QDs/Total injected dose of QDs.

*Tumor section staining and whole-slide imaging.* The frozen tumor blocks were cut into 20-μm thick sections using a cryotome. The sections were pre-treated with Tris-EDTA buffer (pH 8.0) for heat mediated antigen retrieval and blocked with 3% bovine serum albumin (BSA, Servicebio). The primary CD31 antibody (Servicebio, GB13428) was added at a 1:100 dilution and incubated at room temperature overnight. The sections were then rinsed with cold PBS three times and incubated with secondary Alex-Fluor-488 antibody (Servicebio, GB25303) at 1:400 dilution for 50 min. The sections were rinsed with cold PBS again and stained by DAPI for 10 min. An autofluorescence-quencher-containing medium (Servicebio, G1221) was used to mount the sections. The sections were then imaged as whole-slide images using a Nikon's digital eclipse C1 microscope system and analyzed by Qupath software to determine the cell, vessel and QD density [38]. The cell, vessel and QD densities were estimated as the area ratios of DAPI-, Alex-Fluor-488- and QD- positive regions to the whole tumor or the selected regions of interest (ROIs).

*Pre-processing and decomposition of whole-slide images.* The whole-slide images were converted from MRXS format to 8-bit TIFF format by Qupath. The obtained images were padding with zeros and decomposed to patches (512 pixels × 512 pixels) to reduce the computation resource demand. The location of patches was indexed so that they can be recomposed to whole-slide images. Patches were excluded from the training and testing process if their sum value of DAPI and CD31 channels was equal to zero. The intensity of included patches was normalized to the range from -1 to +1.

In detail, the input patches are normalized to [-1,1] by the following equation:

$$input\_patch = 2 * \frac{p}{max(I)} - 1$$

where max(I) is the max value of original image I, and p are patches decomposed from this image. When merging generated patches p_g to a single image, anti-normalization is operated as follow:

$$out\_patch = \frac{(p_g + 1)}{2} * max(I)$$

*GANDA training procedure.* The GANDA is based on generative adversarial networks architecture and consists of a generator network (G) and a discriminator network (D). These two networks compete against each other until they reach Nash equilibrium to optimize the output. An L2-norm penalty on the pixel-wise difference between the generated and the real QD-channel patches is also used to ensure the similarity. The objective function of pixel-wise loss $\ell_{pix}(G)$ is defined as:

$$\ell_{pix}(G) = mean(\|G(x) - z\|_2)$$

The objective functions of adversarial loss $\ell_{adv}(D)$ and $\ell_{adv}(G)$ are defined as:

$$\min_D \ell_{adv}(D) = \frac{1}{2} E_x[(D(x) - 1)^2] + \frac{1}{2} E_z[(D(G(z)))^2]$$

$$\min_G \ell_{adv}(G) = \alpha \ell_{pix}(G) + \beta E_z[(D(G(z)) - 1)^2]$$

where the E refers to the expectation, $x$ refers to the one-channel matrixes (DAPI or CD31) or the two-channel tensor (DAPI + CD31), $z$ refers to the matrixes of QD-channel patches, α and β are the weights of adversarial and pixel-wise loss.

The GANDA was implemented in Keras and TensorFlow framework. The whole-slide images of tumor No. 1-5 were pre-processed and used to train the GANDA. The initial learning rate and the epoch number were set to be 2e-4 and 50 in Adam optimizer. The α and β were set to be 10 and 1. Network training and iterative optimization were performed by an Nvidia RTX 2080 Ti GPU.

*Post-processing and re-composition of generated QD patches.* The trained generator G was used to generate the QDs distribution. The whole-slide images of tumor No. 6 were pre-processed to patches. The obtained one-channel (DAPI or CD31) and two-channel (DAPI+CD31) patches were input to the generator G to generate corresponding QD-channel patches. These patches were recomposed to a whole-slide QD-channel image according to the recorded location index. A merged whole-slide image of tumor No.6 was obtained by merging generated QD channel with real DAPI and CD31 channels on Fiji.

*Quantitative analysis of the generated intratumoral QDs distribution.* The QD density was measured on the generated and real section images of tumor No.6 by Qupath. The linear relationship of measured QD densities between the generated and the real image was analyzed by Graphpad Prism. A distance_transform_edt function from Scipy library was used to calculate the QDs-to-Vessels distances on merged and real whole-slide images of tumor No. 6. The intraclass correlation coefficient (ICC) was calculated by R.

**Supporting Information**
Supporting Information is available from the Wiley Online Library or from the author. The code will be available if requested.

**Acknowledgements**

We greatly appreciate the financial support from the National Natural Science Foundation of China (No. 82022034, 81871420, 81871420, 31930020 and 81920108029) and Jiangsu Province Natural Science Foundation of China (BK20200032). Yuxia Tang and Jiulou Zhang contributed equally to this work.
Received: ((will be filled in by the editorial staff))
Revised: ((will be filled in by the editorial staff))
Published online: ((will be filled in by the editorial staff))

Supporting Information

**GANDA: A Deep Generative Adversarial Network Conditionally Generates Intratumoral Nanoparticles Distribution Pixels-to-Pixels**

*Yuxia Tang#, Jiulou Zhang#, Doudou He, Wenfang Miao, Wei Liu, Yang Li, Guangming Lu\*, Feiyun Wu\*, Shouju Wang\**

**Generator network (G) architecture.** The generator network is based on an FCN architecture. It consists of a contracting path, a ResNet path and an expansive path. Each layer in the contracting path receives the preceding layer output as input with 4D tensors and performs convolution, batch normalization and activation. The stride in the contracting path is set to two for down-sampling and the number of filters is 32, 64 and 128. The ResNet path is a three-layer 128×128×128 convolutional chain linked by residual connection. The expansive path has a systematic structure of the contracting path. The up-sampling is done by using transposed convolution with a stride of two. The number of filters is 128, 64 and 32. The final out is a 512×512×1 image patch. The filter size of the first and last convolutional layer is 7×7. The size of other filters is 3×3.

**Discriminator network (D) architecture.** The discriminator network has a classical convolutional neural network (CNN) architecture. The contracting path has six convolutional layers. The filter size is 4×4 and the stride is set to two. The number of filters is 16, 32, 64, 128, 256 and 512. A fully connected layer follows the contracting path and gives the output as 0 or 1 through a sigmoid operation.

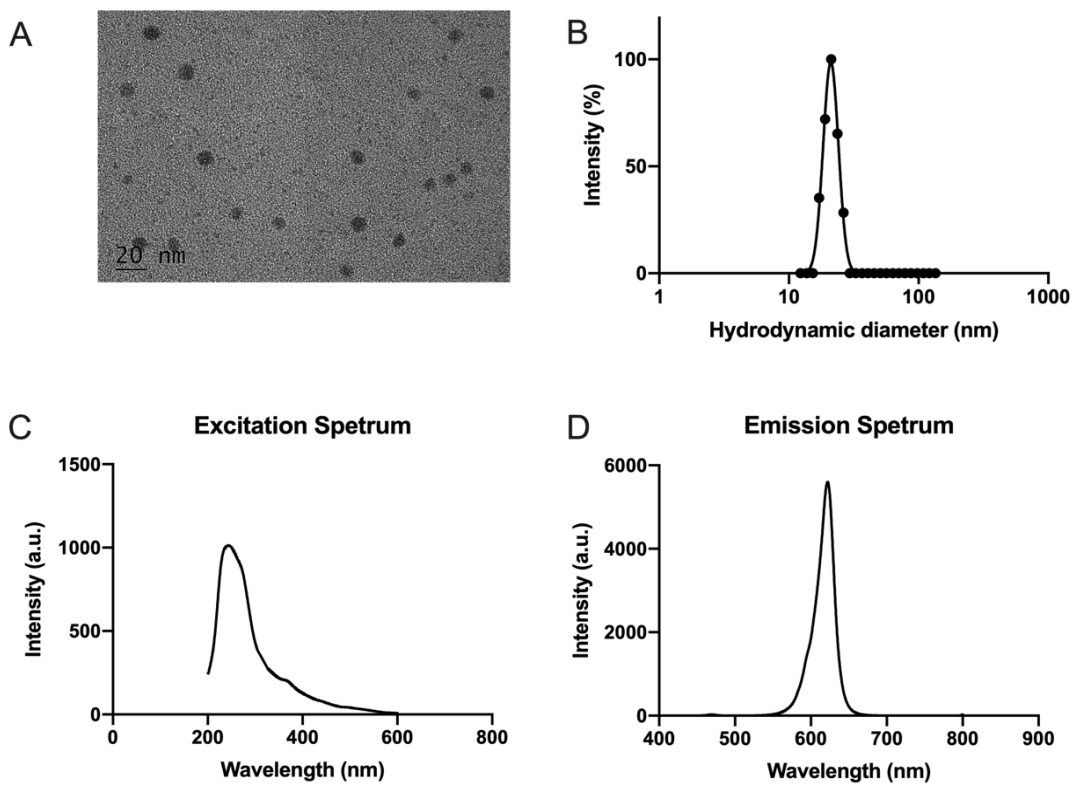

Figure S1. Characteristics of QDs. (A) Transmission electron microscopic image of QDs. Scale bar, 20 nm. (B) Dynamic light scattering intensity distribution of QDs. (C, D) The fluorescent excitation and emission spectra of QDs.

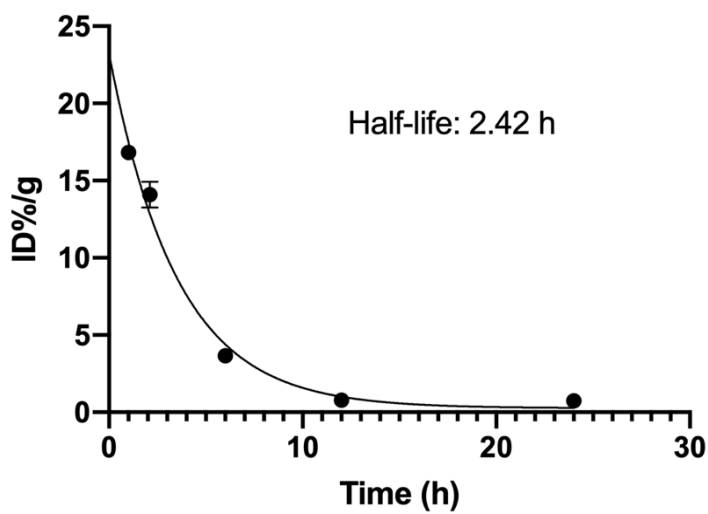

Figure S2. Time course of QDs concentration in the blood samples from mice over 24 h after i.v. injection.

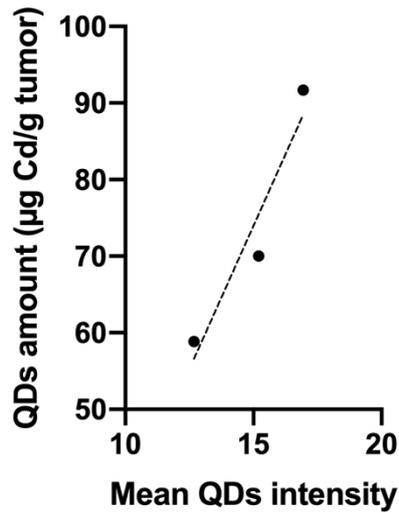

Figure S3. The relationship between the mean QD intratumoral intensity of the section and the total QD amount of the whole tumor.

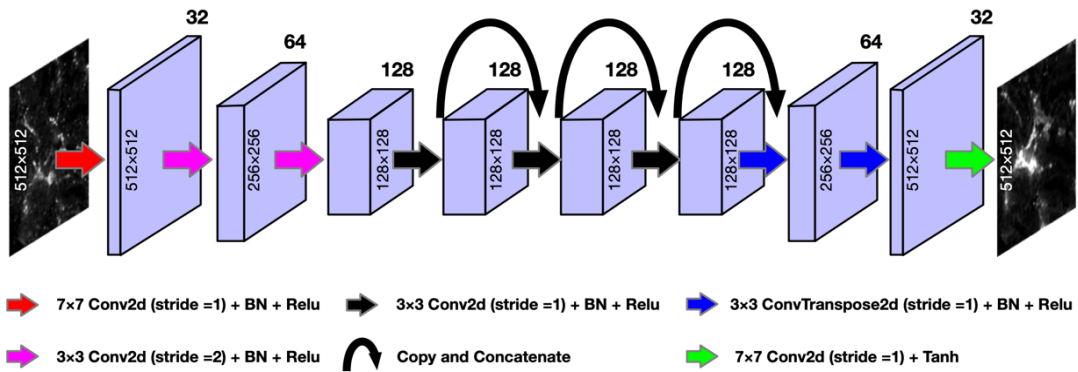

Figure S4. Architecture of generator network. Conv2d refers to 2D convolution; ConvTranspose2d refers to 2D transpose convolution; BN refers to batch normalization.

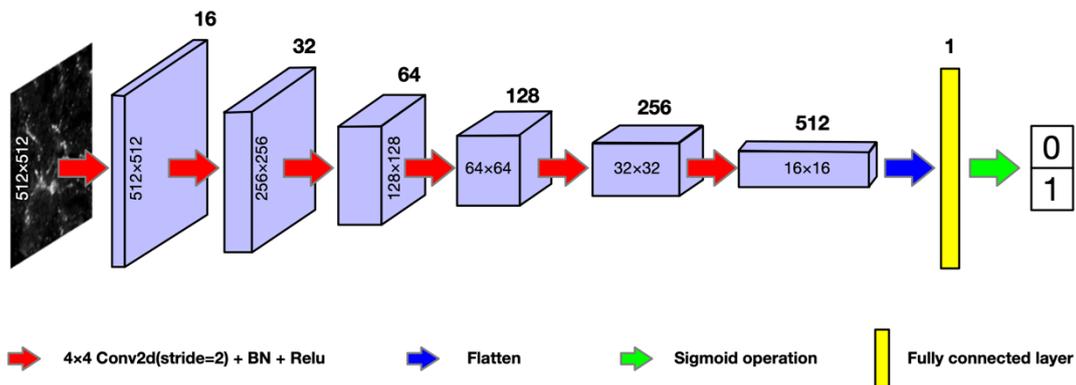

Figure S5. Architecture of discriminator network. Conv2d refers to 2D convolution; BN refers to batch normalization.

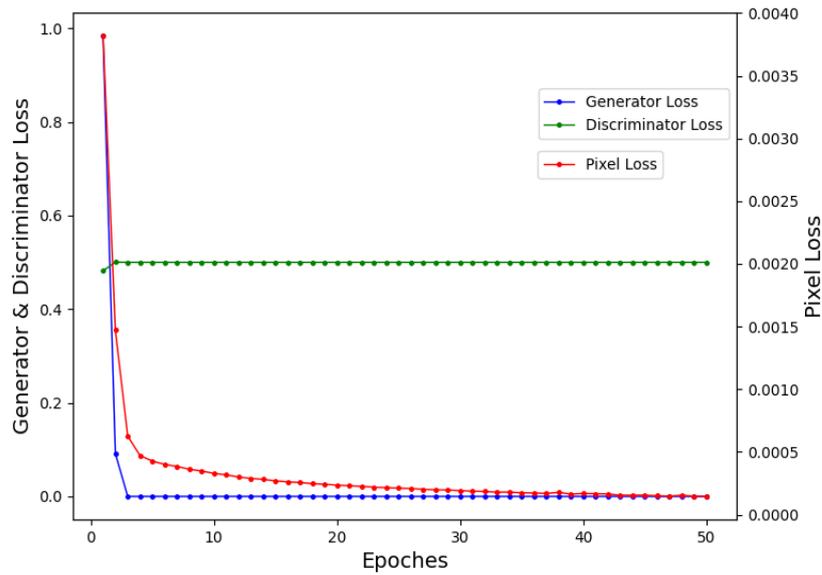

Figure S6. The generator, discriminator, and pixel loss of GANDA trained by DAPI + CD31 during the 50 epoches.

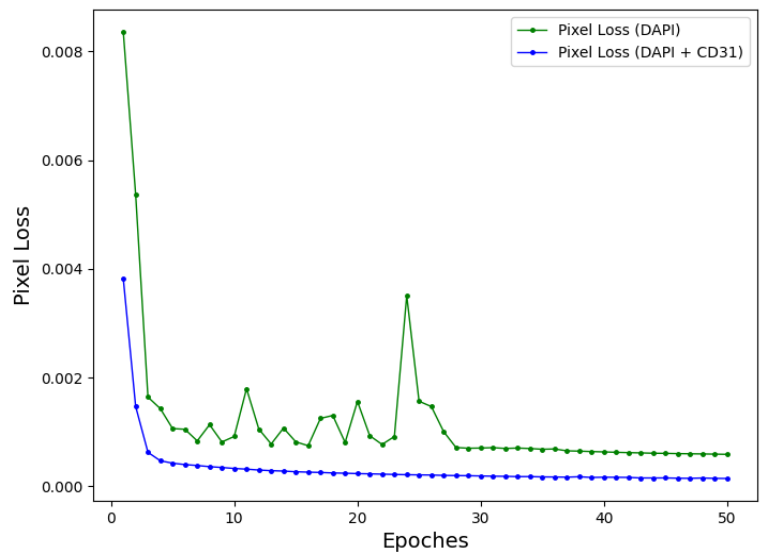

Figure S7. The pixel loss of GANDA trained by DAPI only and DAPI + CD31 during the 50 epoches.

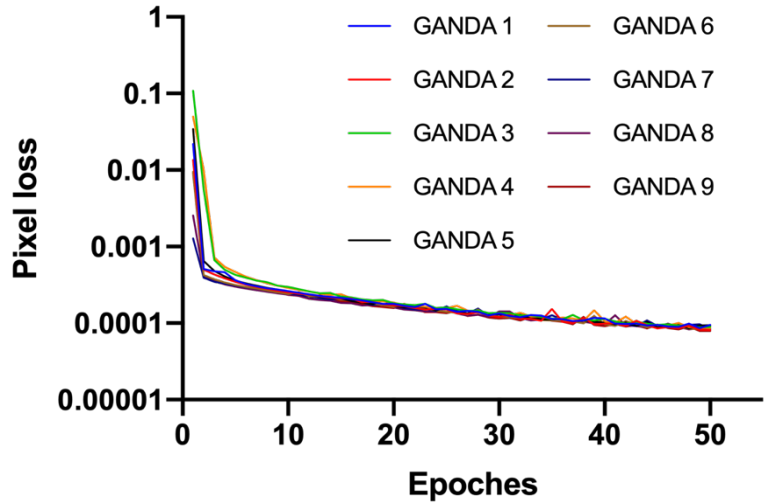

Figure S8. The pixel loss of GANDA (CD31 + DAPI) trained with different initializations during the 50 epoches.

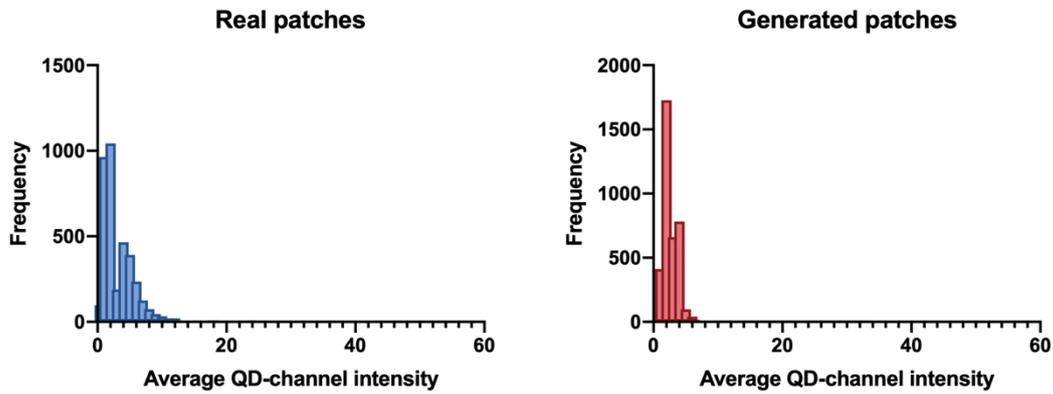

Figure S9. The distribution of average QD-channel intensity of original and generate patches of tumor No. 6.

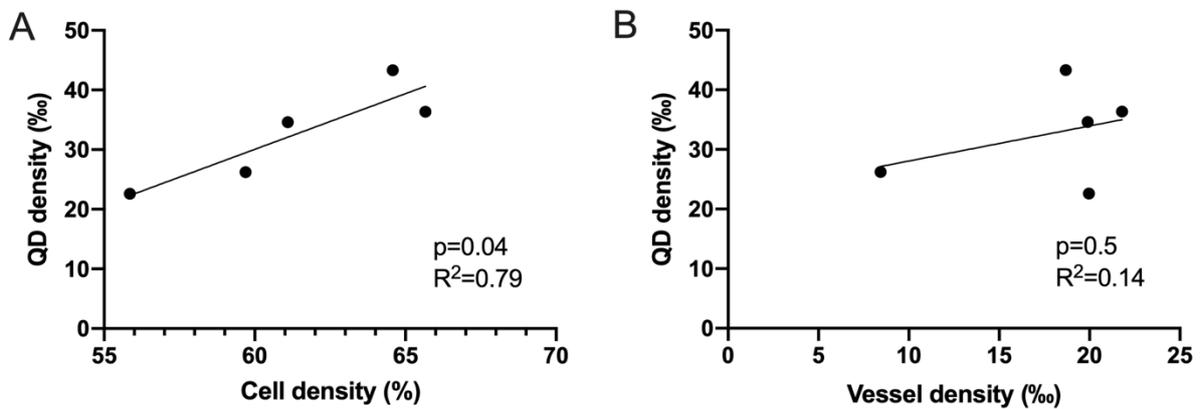

Figure S10. The relationship among cell, vessel and QD density based on the ring-like ROIs of Figure 4F.

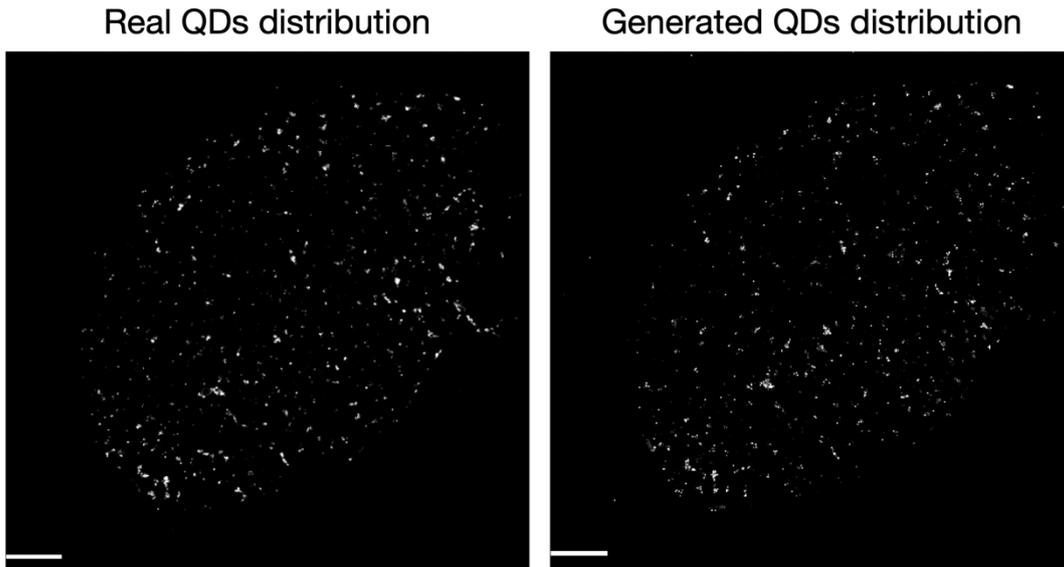

Figure S11. The real and generated QDs distribution of tumor No. 5. The GANDA was trained by the images from Tumor No. 1-4 and 6. Scale bars: 1 mm.

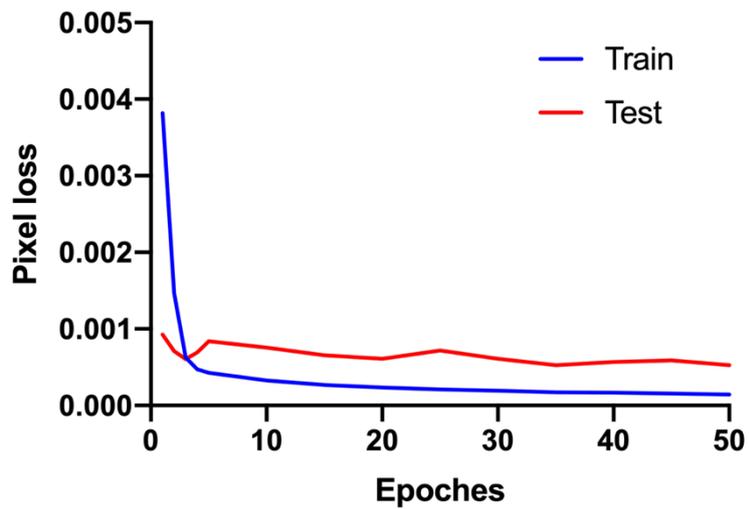

Figure S12. The learning curve in train and test dataset during the 50 epoches of training.

Table S1. The measurement of tumors.

|            | Long axis length (mm) | Short axis length (mm) | Tumor Area (mm$^2$) | Tumor perimeter (mm) |
|------------|------|------|-------|-------|
| Tumor No. 1 | 8.66 | 5.41 | 32.67 | 30.80 |
| Tumor No. 2 | 6.64 | 5.14 | 23.57 | 23.84 |
| Tumor No. 3 | 9.87 | 6.18 | 41.18 | 45.93 |
| Tumor No. 4 | 9.30 | 6.01 | 36.73 | 37.08 |
| Tumor No. 5 | 9.58 | 5.91 | 42.79 | 33.47 |
| Tumor No. 6 | 9.11 | 5.00 | 32.59 | 30.83 |

Table S2. The hydrodynamic diameter and Zeta potential of QDs.

|                          | Hydrodynamic diameter (nm) | Zeta potential (mV) |
|--------------------------|---------------|--------------------|
| 304-CdSe/ZnS-PEG-COOH-620 | 21.04 ± 0.13 | -13.15 ± 1.46 |

Table S3. The summary of extravasation distance analysis in tumor No. 6.

|              | Extravasation distance of real section (μm) | Extravasation distance of synthesized section (μm) |
|--------------|--------|--------|
| 1st quartile | 3.880  | 2.743  |
| median       | 11.640 | 10.974 |
| mean         | 17.193 | 18.694 |
| 3rd quartile | 21.948 | 21.948 |